# Defining, Estimating and Using Credit Term Structures

## Part 1. Consistent Valuation Measures

**Arthur M. Berd**
Lehman Brothers Inc.

**Roy Mashal**
Lehman Brothers Inc.

**Peili Wang**
Lehman Brothers Inc.

*In this three-part series of papers, we argue that the conventional spread measures are not well defined for credit-risky bonds and introduce a set of credit term structures which correct for the biases associated with the strippable cash flow valuation assumption. We demonstrate that the resulting estimates are significantly more robust and remain meaningful even when applied to deeply distressed bonds. We also suggest a new definition of credit bond duration and convexity which remains consistent for distressed bonds and introduce new relative value measures for individual bonds in the context of sector or issuer credit curves, as well as for the basis between cash bonds and credit default swaps (CDS).*

### INTRODUCTION

Most of fixed income valuation and risk methodologies are centered on modeling yield and spread term structures. The main reason for this is that the vast majority of debt instruments exhibit very high correlation of price returns. Therefore the common pricing factors encoded in the yield curve have a high explanatory power. This is especially true for Treasury bonds, where the market is extremely efficient and any deviation of individual bond valuation from the common curve is quickly arbitraged away.

For corporate bonds the common yield curves are much less binding, since the market is substantially less liquid, even for investment grade benchmark issuers. The other driving factors of the valuation of credit-risky bonds are the credit quality of the name, the estimated recovery in case of default, geographic and industry as well as the issuer and security specifics.

The standard market practice in analyzing investment grade credit bonds is to introduce the notion of spread to Treasury or spread to Libor curve. The sector or issuer spread curves are thought to reflect the additional specific information besides the underlying base (Treasury or swaps) yield curve. Among many definitions of spread measures used by practitioners who analyze credit bonds the most robust and consistent one is the Option Adjusted Spread. (OAS, also known as the Z-spread in case of bonds with no embedded options, in the sequel we use these terms interchangeably.)

While the OAS or Z-spread term structures are commonly used to quote and analyze relative value among credit bonds, it is well known that these measures become inadequate for distressed credits for which the market convention reverts to quoting bond prices rather than spreads. This market convention reflects, as we show in later sections, a fundamental flaw in the conventional OAS measure. This methodological breakdown holds for any credits, whether investment grade, or distressed, with the errors being relatively small for investment grade bonds trading near par but growing rapidly with the rise in default risk expectations.

As a simple example let us consider a distressed name with two equal seniority bonds, with 5% coupon, the first matures in 5 years whereas the second in 20 years. As the name is distressed the two bonds are quoted on price, i.e. on the estimated recovery of the name, say 40 cents on the dollar. Using the conventional z-spreads the five year bond may have a Z-spread of over 20,000bps, whereas the 20-year one may have a spread of 900bps. Clearly in this case the Z-spread is not a good measure as both bonds default together and there is actually more probability that the 5-year bond will mature even though it is not reflected at all in its Z-spread.

Thus, in order to correctly assess the relative value between various credit bonds one has to revisit the valuation methodology and introduce new definitions of credit term structures that are consistent with the valuation of bonds for all credit risk levels. As we demonstrate, the resulting relative value and risk measures may differ from conventional ones not only quantitatively but also qualitatively – possibly even reversing the sign of rich/cheap signals.

This article is the first of three papers dedicated to development of consistent credit term structures. The outline of the series is as follows:

- In Part 1 (the present article) we overview the conventional strippable cash flow valuation methodology and explain why it is inadequate for credit-risky bonds. After presenting an alternative framework based on using the term structure of survival probability rather than risky discount function, we introduce a robust estimation methodology based on the exponential splines approximation. We then proceed to define a variety of useful credit term structures for issuers or credit sectors which are consistent with the survival-based valuation of credit bonds, replacing the conventional nominal spread, OAS or Z-spread term structures as the determinants of the relative value across issuers or sectors. Finally, we complement these issuer- and sector-specific term structures with bond-specific valuation measures.
- In Part 2 we redefine bond-specific risk measures and analyze their differences from the conventional measures, in particular emphasizing important variations in the duration and convexity estimates for credit bonds. We demonstrate that these differences produce a material impact on construction of long-short portfolios of credit bonds.
- In Part 3 we apply the newly developed notions such as the bond-implied CDS term structure to the analysis of the relative value between cash credit bonds and credit default swaps (CDS). We demonstrate the complementarity between cash bonds and forward CDS and construct the corresponding static hedging/replication strategies. The resulting relative value measures for basis trading exhibit a non-trivial dependence on the term structure of the underlying interest rate and the issuer credit curve.

The Parts 2 and 3 of this series will appear in the subsequent issues of the Financial Analyst Journal.

## CREDIT BONDS AND THE STRIPPABLE CASH FLOW VALUATION METHODOLOGY

In this Section we re-examine critically the standard lore of credit bond valuation methodologies in order to understand better their breakdown in case of distressed bonds which we mentioned in the Introduction.

### The strippable cash flow valuation methodology

The main assumption of the strippable cash flow valuation methodology is that fixed-coupon debt instruments can be priced as a portfolio of $N$ individual cash flows $CF(t_i)$ using the discount function $Z(t_i)$, which coincides with the present value of a zero coupon bond maturing at some future time $t_i$ as observed today. The present value of all cash flows, including coupon and principal payments, is then given by a simple sum:

[1] $$PV_{bond} = \sum_{i=1}^{N} CF(t_i) \cdot Z(t_i)$$

Such an assumption is clearly valid for default risk-free instruments such as U.S. Treasury bonds. The discount function $Z(t)$ itself is commonly represented in terms of yield curve term structure. One of the most convenient parameterization is given by the term structure of zero-coupon bond yields $y_z(t)$ in a continuous compounding convention:

[2] $$Z(T) = e^{-y_z(T) \cdot T}$$

The strippable cash flow methodology is commonly extended to credit-risky bonds by assuming that they can be priced using a "risky yield curve" $Y_z(t)$ which is assumed to differ from the underlying base yield curve by an amount often referred as the Z-spread, $S_z(t)$ (for an exact definitions of the Z-spread as well as other conventional credit spread measures see O'Kane and Sen [2004]):

[3] $$Y_z(t) = y_z(t) + S_z(t)$$

As a consequence of this definition one can immediately see that the "risky" discount function is a product of the base discount function and the spread discount function:

[4] $$Z_{risky}(T) = Z_{base}(T) \cdot Z_{spread}(T)$$

where the spread discount function is defined as:

[5] $$Z_{spread}(T) = e^{-S_z(T) \cdot T}$$

Thus, in the conventional strippable cash flow methodology the fundamental pricing equation for a credit risky bond reads as follows:

[6] $$PV_{bond} = \sum_{i=1}^{N} CF(t_i) \cdot Z_{base}(t_i) \cdot Z_{spread}(t_i)$$

While the considerations above refer to a specific choice of spread measure (namely the Z-spread) and a specific choice of compounding convention (namely continuous), the final relationship has actually a much wider applicability within the conventional framework and can actually be considered a definition of this framework:

*The present value of a contractually-fixed cash flow security under a strippable cash flows valuation framework is equal to the sum of the present values of the individual cash flows.*

In other words, the strippable cash flows framework hinges on the ability to represent a fixed income security as a portfolio of individual cash flows. Whether or not such a representation is possible, and what discount function applies if it is, depends on the realities of the market as we discuss next.

## Credit bond cash flows reconsidered

The root of the problem with the conventional strippable cash flow methodology as applied to credit-risky bonds is that **credit bonds do not have fixed cash flows**. Indeed, the main (some would say the only) difference between a credit risky bond and a credit risk-free one is precisely the possibility that the issuer might default and will not honor the contractual cash flows of the bond. In this event, even if the contractual cash flows were fixed, the realized cash flows may be very different from the promised ones and highly variable.

Once we realize this fundamental fact, it becomes clear that the validity of the representation of a credit bond as a portfolio of cash flows critically depends on our assumption of the cash flows in case of default. In reality, once an issuer defaults it enters into an often protracted bankruptcy proceedings during which various creditors including bondholders, bank lenders, and those with trading and other claims on the assets of the company settle with the trustees of the company and the bankruptcy court judge the priority of payments and manner in which those payments are to be obtained.

Of course, modeling such an idiosyncratic process is hopelessly beyond our reach. Fortunately, however, this is not necessary. Once an issuer defaults or declares bankruptcy its bonds trade in a fairly efficient distressed market and quickly settle at what the investors expect is the fair value of the possible recovery.

The efficiency of the distressed market and the accuracy and speed with which it zooms in on the recovery value is particularly high in the U.S., as evidenced by many studies (see Gupton and Stein [2002] and references therein). Assuming that the price of the bond immediately after default represents the fair value of the subsequent ultimate recovery cash flows, we can simply take that price as the single post-recovery cash flow which substitutes all remaining contractual cash flows of the bond.

The market practice is to use the price approximately one month after the credit event to allow for a period during which investors find out the extent of the issuer's outstanding liabilities and remaining assets. This practice is also consistent with the conventions of the credit derivatives market, where the recovery value for purposes of cash settlement of CDS is obtained by a dealer poll within approximately one month after the credit event.

From a modeling perspective, using the recovery value after default as a replacement cash flow scenario is a well established approach. However, the specific assumptions about the recovery value itself differ among both academics and practitioners. We will discuss these in detail in the next section as we develop a valuation framework for credit bonds. But first we would like to explore a few more general aspects of credit bond valuation that will set the stage for an intuitive understanding of the formal methodology we present later in the article.

### How important is the uncertainty of the cash flows?

As we have already mentioned, the uncertainty of cash flows is the main difference between credit risky and default free bonds. It is the possibility of a non-contractual change in the cash flows that drives the pricing of these bonds. Therefore, the right emphasis is not whether the uncertainty of the cash flows is important or not – as it is always important – but rather whether the mistakes one makes when not taking it explicitly into account are small enough to be ignored in practice.

The relevance of the uncertain cash flows of credit-risky bonds depends on the likelihood of the scenarios under which the fixed contractual cash flows may not be honored. In other words, it depends on the level of default probability. As we will find out shortly, what one has to consider here is not the real-world (forecasted) default probability, but the so-called implied (or breakeven) default probability.

One of the most striking examples of the mis-characterization of credit bonds by the conventional spread measures is the often cited steeply inverted spread curves for distressed bonds. One frequently hears explanations for this phenomenon based on the belief that the near-term risks in distressed issuers are greater than the longer term ones which is supposedly the reason why the near-term spreads are higher than the long maturity ones. However, upon closer examination one can see that the inverted spread curve is largely an "optical" phenomenon due to a chosen definition of the spread measure such as the Z-spread rather than a reflection of the inherent risks and returns of the issuer's securities.

Indeed, the more important phenomenon in reality is that once the perceived credit risk of an issuer become high enough, the market begins to price the default scenario. In particular, investors recognize what is known in the distressed debt markets as the "acceleration of debt" clause in case of default. The legal covenants on most traded bonds are such that, regardless of the initial maturity of the bond, if the issuer defaults on any of its debt obligations, all of the outstanding debt becomes due immediately. This is an important market practice designed to make sure that, in case of bankruptcy, investor interests can be pooled together by their seniority class for the purposes of the bankruptcy process. As a result, both short and long maturity bonds begin trading at similar dollar prices – leading to a "flat" term structure of *prices*.

Let us now analyze how this translates into a term structure of Z-spreads. In the conventional spread-discount-function based methodology, one can explain an $80 price of a 20-year bond with a spread of 500bp. However, to explain an $80 price for a 5-year bond, one would need to raise the spread to very large levels in order to achieve the required discounting effect. The resulting term structure of spreads is, of course, downward sloping, or inverted. The inversion of the spread curve is due to the bonds trading on price, i.e. on their recovery while the Z-spread methodology does not take recoveries at all into account! In the survival-based methodology, which we will describe later in this paper, the low prices of bonds are explained by high default rates, which need not have an inverted term structure. A flat or even an upward sloping term structure of default rates can lead to an inverted Z-spread term structure if the level of the default rates is high enough. This is not to say that the near-term perceived credit risks are never higher than the longer term ones – just that one cannot make such a conclusion based on the inverted Z-spread term structure alone.

Consider for example the credit term structure of Ford Motor Co. as of 12/31/2002 shown in Figure 1. The default hazard rate were fitted using the survival-based methodology developed later in this paper is upward sloping with a hump at 15 year range. However, the fitted Z-spread curve using the conventional methodology which does not take the potential variability of cash flows is clearly inverted.

**Figure 1. Fitted hazard rate and Z-spread term structures, Ford as of 12/31/02**

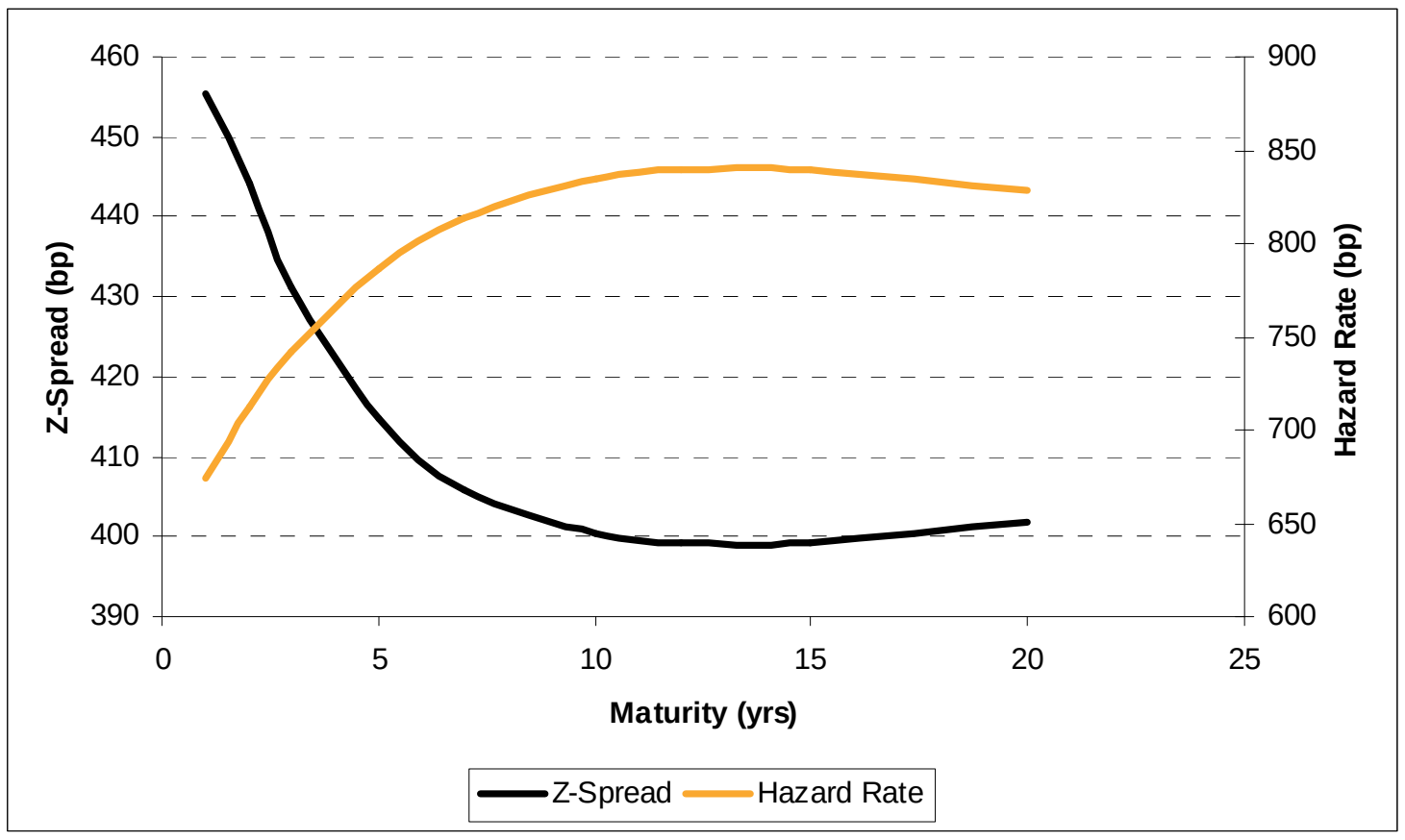

**Figure 2. CDS, Z-spread and BCDS term structures, Ford as of 12/31/02 (source: Lehman Brothers, Mark-it Partners)**

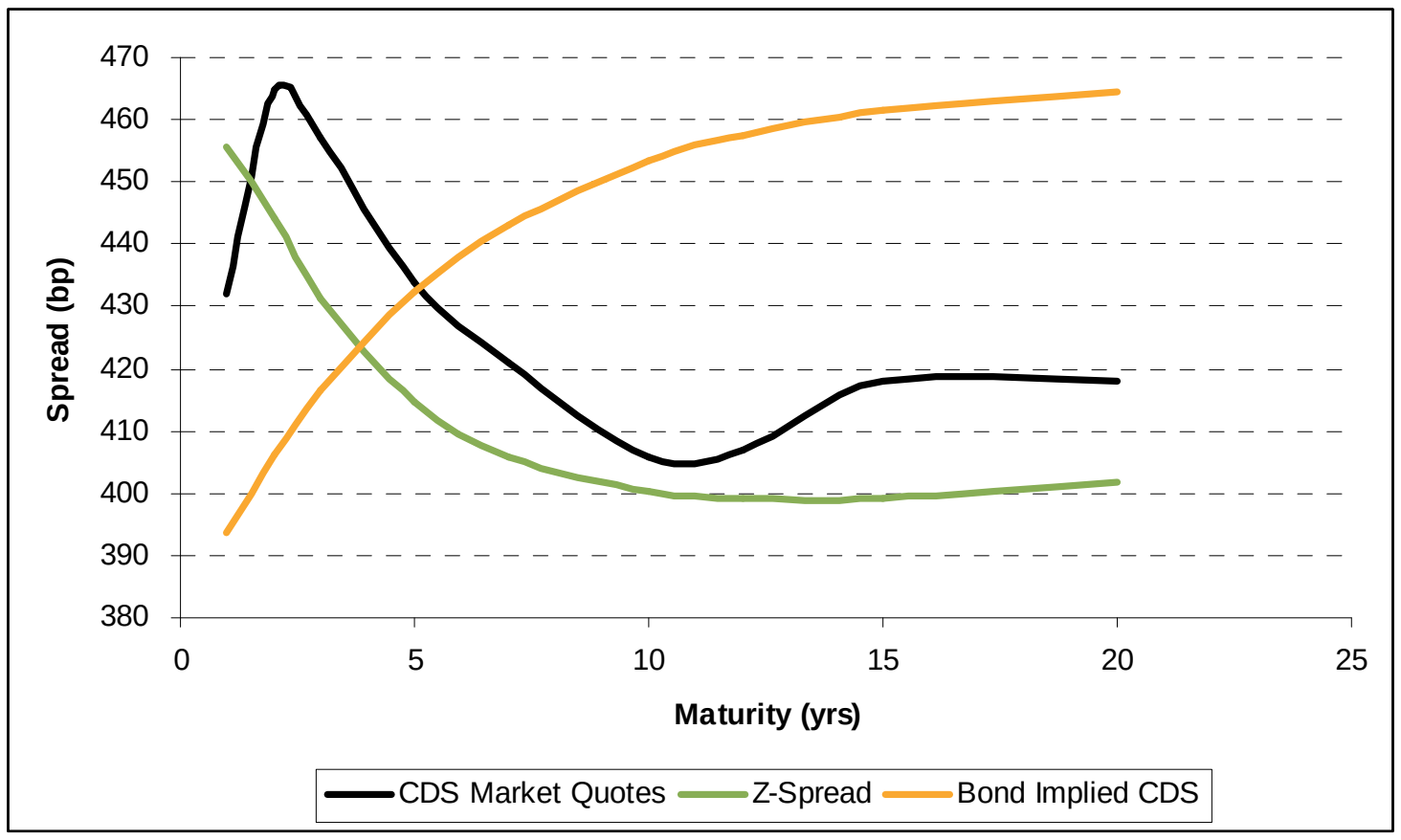

Figure 2 shows the Z-spread, Credit Default Swap (CDS), and the bond-implied CDS (BCDS) term structures of Ford for the same date. From the figure it is clear that investors who looked at the Z-spread term structure and compared it to the cost of protection available via the credit default swap (CDS) market, would have been misled to think that there was a small positive basis between the two, i.e. that the CDS traded wider by roughly 20 bp than bonds across most maturities, with the shape of the curve following closely the shape of the Z-spread curve with the exception of less than 2 year maturities. In fact, if one had used the methodology presented in this paper and derived the BCDS term structure and compared it with the market quotes, one would see a very different picture – the bond market traded more than 50 bp tighter than CDS at short maturities and more than 50 bp wider at maturities greater than 5 years, if measured on an apples-to-apples basis.

Both the bonds and the CDS of Ford Motor Credit are among the most liquid and widely traded instruments in the U.S. market and these differences are clearly not a reflection of market inefficiency but rather of the difference between the methodologies. Only for the very short term exposures, where investors are forced to focus critically on the likelihood of the default scenarios and the fair value of the protection with a full account for cash flow outcomes do we see the CDS market diverging in shape from the cash market's "optically distorted" perception of the credit spreads. We define the BCDS term structure later in this paper, and address the CDS-Bond basis in a much greater detail in Part 3 of this series.

## SURVIVAL-BASED MODELING OF CREDIT-RISKY BONDS

In this Section we consider the specific modeling assumptions and derive the empirical estimation procedures for the survival-based modeling of credit risky bonds.

### Reduced-form modelling and recovery assumptions

The standard methodology for valuation of credit-risky bonds is the reduced-form approach (see the pioneering works by Litterman and Iben [1991], Jarrow and Turnbull [1995], Jarrow, Lando and Turnbull [1997] and Duffie and Singleton [1999] as well as the textbooks by Duffie and Singleton [2003] and Schonbucher [2003] for detailed discussions and many more references). The key assumption in these models is what will be the market value of a bond just after default. In other words, one must make an assumption on what is the expected recovery given default (or alternatively what is the loss given default). There are three main conventions regarding this assumption:

- **Fractional recovery of Treasury (FRT).** In this approach, following Jarrow and Turnbull (1995), one assumes that upon default a bond is valued at a given fraction to the hypothetical present value of its remaining cash flows, discounted at the riskless rate.
- **Fractional recovery of market value (FRMV).** Following Duffie and Singleton (1999), one assumes in this approach that upon default a bond loses a given fraction of its value just prior to default.
- **Fractional recovery of par (FRP).** Under this assumption, a bond recovers a given fraction of its face value upon default, regardless of the remaining cash flows. A possible extension is that the bond additionally recovers a (possibly different) fraction of the current accrued interest.

Both FRMV and FRT assumptions lead to very convenient closed form solutions for pricing defaultable bonds as well as derivatives whose underlying securities are defaultable bonds. In both approaches one can find an equivalent "risky" yield curve which can be used for discounting the promised cash flows of defaultable bonds and proceed with valuation in essentially the same fashion as if the bond was riskless – the only difference is the change in the discounting function. As a result, either of these approaches works quite well for credit-risky bonds that trade not too far from their par values (see Jarrow and Turnbull [2000], Duffie and Singleton [2003] and Schonbucher [2003] for related discussions).

The main drawback of both of these approaches is that they do not correspond well to the market behaviour when bonds trade at substantial price discounts. Namely, the FRMV assumption fails to recognize the fact that the market begins to discount the future recovery when the bonds are very risky. In other words, the bonds are already trading "to recovery" just prior to default, therefore there is relatively little additional loss when the actual event of default takes place. The FRT assumption, on the other hand, does not recognize the acceleration of debt, the importance of which we highlighted in the previous section.

Of course, both FRMV and FRT approaches can be adjusted to conform to market behaviour by generalizing the expected recovery from a constant to a variable dependent on the current price of the bonds. However, such a generalization would invalidate the closed-form expressions for risky yields and negate the main advantage of these recovery assumptions. In fact, we think that what is normally considered to be an advantage of the FRMV and FRT recovery assumptions is actually a deficiency. Namely, the possibility of a strippable cash flow valuation under these assumptions with the present value of a credit bond being a simple sum of the present values of contractual cash flows is in contradiction with our understanding that all credit bonds, regardless of their contractual structure, have an embedded option to default and therefore they simply cannot be thought of as just a portfolio of coupon and principal cash flows – irrespective of whether the inherent option to default is exercised by a limited liability issuer rationally or is triggered by exogenous factors.

The only limiting case when the FRMV and FRT assumptions are in agreement with the acceleration of debt and equal priority recovery is when the expected recovery value is precisely equal to zero. We denote this as the zero recovery (ZR) assumption. As we will see in the subsequent sections, these three cases (FRMV, FRT and ZR recovery assumptions) are the only cases when the valuation of a bond as a portfolio of individual contractual cash flows remains valid despite the possibility of a default scenario. This is simply due to the fact that under these assumptions one does not introduce any new cash flow values which were not already present in bond's contractual description. Therefore, when calculating the present value of a credit bond, one can combine the riskless discount function, the survival probability, and the assumed recovery fraction into a "risky discount function" which can then be applied to the contractual cash flows (see Duffie, Schroder and Skiadas [1996] for a discussion of conditions under which the valuation of defaultable securities can be performed by applying a risky stochastic discount process to their default-free payoff stream).

In contrast, the fractional recovery of par (FRP) assumption is fully consistent with the market dynamics and behaviour, and can explain some of the salient features of distressed credit pricing in a very intuitive manner as discussed in the previous section. Given the recent experience of the credit markets in 2001-2002 with a large number of fallen angels trading at deep discounts, we feel that it is worthwhile to explore the FRP assumption in greater detail and derive the corresponding relative value measures.

Furthermore, with the advent of the credit default swap (CDS) market, the fractional recovery of par assumption has taken on a new importance. The market convention in modeling CDS spreads follows the FRP assumption and therefore the discrepancy between the CDS and conventional bond pricing models can be large in case of bonds that are trading with very wide spreads, as we demonstrated earlier in Figure 2. A quantitatively consistent comparison between non-par bonds and par-equivalent CDS which is necessary for the analysis of CDS-Cash basis investment strategies that have become very popular in the past several years, also requires a consistent recovery assumption across both types of securities.

Despite the widespread use of alternative recovery assumptions by practitioners and academics, there are only a handful of studies which examine their importance for pricing of standard credit bonds and CDS. Finkelstein (1999) has pointed out the importance of the correct choice of the recovery assumption for the estimation of the term structure of default probabilities when fitted to observed CDS spreads, and the fact that the strippable valuation of credit bonds becomes impossible under the FRP assumption.

Duffie (1998) has explored the pricing of default-risky securities with fractional recovery of par and derived the risk-neutral dynamics of hazard rates. In particular, he obtained a generic result relating the instantaneous implied hazard rate to the short spread via the widely used credit triangle formula $h = S/(1-R)$. The important question, however, is what is the meaning of the spread used in this relationship. Under the assumptions in Duffie (1998), this is the spread of a zero-coupon principal strip – an asset that is not actually observed in marketplace. We will discuss later in this paper that, while such a characterization of credit spreads is possible, it is not very useful in practice, and that an alternative definition of spreads is preferable that refers to either full coupon-bearing bonds or CDS.

Bakshi, Madan and Zhang (2004) have specifically focused on the implications of the recovery assumptions for pricing of credit risk. Having developed a valuation framework for defaultable debt pricing under all three recovery assumptions (FRMV, FRT and FRP), they have concluded from the comparison with time series of 25 liquid BBB-rated bullet bonds that the FRT assumption fits the bond prices best. While we agree with their methodology in general, we believe that in this case the "market is wrong" for a variety of legacy reasons discussed in the previous section, and one must insist on a better FRP model *despite* the empirical evidence from the investment grade bond prices. Our estimates, based on 10 years of monthly prices for 5000+ senior unsecured bonds across all rating categories, suggest that the FRP assumption allows for a more robust fit across a larger sample.

Finally, in an important recent empirical study Guha (2002) examined the realized recoveries of U.S.-based issuers and concluded that the FRP assumption is strongly favored by the data in comparison to the FRMV or FRT assumptions. In particular, he has shown that the vast majority of defaulted bonds of the same issuer and seniority are valued equally or within one dollar, irrespective of their remaining time to maturity.

### Survival-based pricing of bonds with fractional recovery of par

Consider a credit-risky bond with maturity *T* that pays fixed cash flows with specified frequency (usually annual or semi-annual). According to the fractional recovery of par assumption, the present value of such a bond is given by the *expected discounted* future cash flows, including the scenarios when it defaults and recovers a fraction of the face value and possibly of the accrued interest, discounted at the risk-free (base) rates. By writing explicitly the scenarios of survival and default, we obtain the following pricing relationship at time *t*:

$$\begin{aligned} PV(t) &= \sum_{t_i > t}^{t_N = T} \left( CF^{prin}(t_i) + CF^{\text{int}}(t_i) \right) \cdot \mathrm{E}_t \left\{ Z_{t_i} \cdot I_{\{t_i < \tau\}} \right\} \\ &+ \int_t^T \mathrm{E}_t \left\{ \left( R_p \cdot F^{prin}(\tau) + R_c \cdot A^{\text{int}}(\tau) \right) \cdot Z_\tau \cdot I_{\{u < \tau \le u + du\}} \right\} \end{aligned} \quad [7]$$

The variable $\tau$ denotes the (random) default time, $I_{\{X\}}$ denotes an indicator function for a random event *X*, $Z_u$ is the (random) credit risk-free discount factor, and $\mathrm{E}_t\{\bullet\}$ denotes the expectation under the risk-neutral measure at time $t$.

The first sum corresponds to scenarios in which the bond survives until the corresponding payment dates without default. The total cash flow at each date is defined as the sum of principal $CF^{prin}(t_i)$, and interest $CF^{\text{int}}(t_i)$, payments. The integral corresponds to the recovery cashflows that result from a default event occurring in a small time interval $[u, u + du]$, with the bond recovering a fraction $R_p$ of the outstanding principal face value $F^{prin}(\tau)$ plus a (possibly different) fraction $R_c$ of the interest accrued $A^{\text{int}}(\tau)$.

Assuming the independence of default times, recovery rates[1] and interest rates[2], one can express the risk-neutral expectations in eq. [7] as products of separate factors encoding the term structures of (non-random) base discount function, survival probability and conditional default probability, respectively:

$$\begin{cases} Z_{base}(t,u) &= \mathrm{E}_t\{Z_u\} \\ Q(t,u) &= \mathrm{E}_t\{I_{\{u<\tau\}}\} \\ D(t,u) &= \dfrac{d}{du}\mathrm{E}_t\{I_{\{u<\tau<u+du\}}\} = -\dfrac{d}{du}Q(t,u) \end{cases} \quad [8]$$

Many practitioners simply use a version of the equation [7] assuming that the recovery cash flows occurs on the next coupon day $\tau = t_i$, given a default at any time within the previous coupon payment period $[t_{i-1}, t_i]$. As a possible support for this assumption, one might argue that the inability to meet company's obligations is more likely to be revealed on a payment date than at any time prior to that when no payments are due, regardless of when the insolvency becomes inevitable. Of course, this argument becomes much less effective if the company has other obligations besides the bond under consideration. Still, to simplify the implementation we will follow this approach in the empirical section of this paper.

---

*[1] Altman et. al. [2003] discuss the empirical evidence for a negative correlation between recovery rates and default rates which may have a significant impact on the pricing of credit bonds.*

*[2] Jarrow (2001) and Jarrow and Yildirim (2002) have considered the pricing of credit-risky securities under correlated interest and hazard rates. However, their model presumes the validity of the strippable cash flow valuation of credit bonds, and is therefore not applicable under the FRP assumption adopted here. Having said this, the effect of a an observed (negative) correlation between the hazard rates and interest rates can be significant.*

We also ignore the recovery of the accrued interest and set $R_c = 0$. In general, we do not believe that the market prices the coupon recovery to any significant degree and we do not know of any reasonable way to estimate this parameter. Any potential inaccuracy caused by these assumptions is subsumed by the large uncertainty about the level of the principal recovery, which is much more important.

For the case of fixed-coupon bullet bonds with coupon frequency *f* (e.g. semi-annual *f=2*) and no recovery of the accrued coupon, we have a simplified version of the pricing equation:

[9]
$$\begin{aligned} PV &= Z_{base}(t_N) \cdot Q(t_N) + \frac{C}{f} \cdot \sum_{i=1}^{N} Z_{base}(t_i) \cdot Q(t_i) \\ &+ R_p \cdot \sum_{i=1}^{N} Z_{base}(t_i) \cdot D(t_{i-1}, t_i) \end{aligned}$$

We have dropped for simplicity the argument denoting the valuation time *t*. The probability $D(t_{i-1}, t_i)$ that the default will occur within the time interval $[t_{i-1}, t_i]$, conditional on surviving until the beginning of this interval, is related to the survival probability in a simple, reflecting the conservation of total probability (also seen from definition in [8]):

[10]
$$D(t_{i-1}, t_i) = Q(t_{i-1}) - Q(t_i)$$

Substituting this definition, taking into account that the survival probability for the current time is $Q(t_0) \equiv Q(0) = 1$, and rearranging the terms in equation [9], we derive the following survival-based pricing formula for a fixed-coupon credit-risky bond:

[11]
$$\begin{aligned} PV &= \sum_{i=1}^{N-1} Q(t_i) \cdot \left[ \frac{C}{f} \cdot Z_{base}(t_i) - R_p \cdot \left( Z_{base}(t_i) - Z_{base}(t_{i+1}) \right) \right] \\ &+ Q(t_N) \cdot Z_{base}(t_N) \cdot \left[ \frac{C}{f} + 1 - R_p \right] + R_p \cdot Z_{base}(t_1) \end{aligned}$$

One can see quite clearly from this expression that under the fractional recovery of par (FRP) assumption the present value of the coupon-bearing credit bond does not reduce to a simple sum of contractual cash flow present values using any risky discount function. An obvious exception to this is the case of zero recovery assumption, under which the survival probability plays the role of a risky discount function.

## ESTIMATING SURVIVAL RATES WITH EXPONENTIAL SPLINES

Having derived the pricing relationship in the survival-based approach, we are now ready to estimate the implied survival probability term structure directly from bond prices. The premise of our approach is that the survival probability is generally an exponentially decaying function of maturity, perhaps with a varying decay rate. This assumption is generally valid in Poisson models of exogenous default, where the default hazard rate is known but the exact timing of the default event is unpredictable. This is the same assumption made by all versions of reduced-form models regardless of the recovery assumptions discussed earlier. Notably, this assumption differs from the Merton-style structural models of credit risk (Merton [1974]) where the timing of default becomes gradually more predictable as the assets of the firm fall towards the default threshold.

When it comes to the estimation of term structures based on a large number of off-the-run bonds across a wide range of maturities, most approaches based on yield or spread fitting are not adequate because they lead to a non-linear dependence of the objective function on the fit parameters. The most important aspect of this problem is the large number of securities to be fitted which makes a precise fit of all prices impractical (or not robust) and therefore creates a need for a clear estimate of the accuracy of the fit. After all one must know whether a given bond trading above or below the fitted curve represents a genuine rich/cheap signal or whether this mismatch is within the model's error range. Without such estimate relative value trading based on fitted curves would not be possible.

Vasicek and Hong (1982) (see also Shea [1985]) suggested a solution to this dilemma, which has become a de-facto industry standard for off-the-run Treasury and agency curve estimation[3]. They noted that the above problem is best interpreted as a cross-sectional regression. As such, it would be best if the explanatory factors in this regression were linearly related to the observable prices, because this would lead to a (generalized) linear regression. Realizing further that the quantity which is linearly related to bond prices is the discount function, they proposed to estimate the term structure of risk-free discount function itself rather than the term structure of yields. Finally, they argued that the simplest discount function is exponentially decreasing with a constant rate, and concluded that one must use some linear combination of exponential functions to best approximate the shape of realistic discount functions. We review the definition of exponential splines in the Appendix.

In the case of credit risky bonds a similar logic also applies, except that one has to think about the survival probabilities rather than discount function, because it is the survival probabilities that appear linearly in the bond pricing equation [11]. When the hazard rate is constant the survival probability term structure is exactly exponential. Therefore, it is indeed well suited for approximation by exponential splines [39]:

[12] $$Q(t) = \sum_{k=1}^{K_{surv}} \beta_k^{surv} \cdot SSpline_k\left(t \middle| \alpha^{surv}\right)$$

where the spline factors $SSpline_k\left(t \middle| \alpha^{surv}\right)$ depend on the tenor $t$ and on the long-term decay factor $\alpha^{surv}$, which in this case has the meaning of the generic long-term hazard rate.

Assuming that we have already estimated the base discount function, and substituting the spline equation [12] into pricing equation [11] we obtain a cross-sectional regression setting for direct estimation of the survival probability term structure from observable bond prices:

[13] $$V_q = \sum_{k=1}^{K_{surv}} \beta_k^{surv} \cdot U_q^k + \varepsilon_q$$

which in matrix notations looks like:

[14] $$\begin{bmatrix} V_1 \\ \vdots \\ V_q \\ \vdots \end{bmatrix} = \begin{bmatrix} U_1^1 & U_1^2 & U_1^3 \\ \vdots & \vdots & \vdots \\ U_q^1 & U_q^2 & U_q^3 \\ \vdots & \vdots & \vdots \end{bmatrix} \cdot \begin{bmatrix} \beta_1^{surv} \\ \beta_2^{surv} \\ \beta_3^{surv} \end{bmatrix} + \begin{bmatrix} \varepsilon_1 \\ \vdots \\ \varepsilon_q \\ \vdots \end{bmatrix}$$

Here we introduced the explanatory variables $U_q^k$ and the adjusted present value $V_q^{surv}$ as:

*[3] We note that for term structures based on a few well defined benchmark securities such as the on-the-run Treasuries curve or the swaps curve alternative estimation techniques including bootstrapping and cubic spline smoothing may be more appropriate (see for example Adams and van Deventer [1994] and Zhou [2000]).*

[15]
$$U_q^k = \sum_{i=1}^{N-1} SSpline_k\left(t_i^q\right)\cdot\left[\frac{C_q}{f_q}\cdot Z_{base}(t_i^q) - R_p\cdot\left(Z_{base}(t_i^q) - Z_{base}(t_{i+1}^q)\right)\right] + SSpline_k\left(t_N^q\right)\cdot\left(\frac{C_q}{f_q} + \left(1-R_p\right)\right)\cdot Z_{base}(t_N^q)$$

[16]
$$V_q = PV_q - R_p\cdot Z_{base}(t_1^q)$$

We have found empirically that it is often sufficient to retain only the first three factors for estimating the survival probability. Thus there are no knot-factors in our implementation of this approach. The first three coefficients of the spline expansion must satisfy an equality constraint, because the survival probability must be exactly equal to 1 when the time horizon is equal to zero.

[17]
$$\sum_{k=1}^{3}\beta_k^{surv} = 1$$

In addition to the equality constraint, we also impose inequality constraints at various maturities $T_c$ to make sure that the survival probability is strictly decreasing, and consequently the hazard rate is strictly positive. Their functional form is:

[18]
$$\sum_{k=1}^{3}\beta_k^{surv}\cdot k\cdot e^{-k\cdot\alpha^{surv}\cdot T_c} > 0$$

In addition, we impose a single constraint at the long end of the curve to make sure that the survival probability itself is positive.

[19]
$$\sum_{k=1}^{3}\beta_k^{surv}\cdot e^{-k\cdot\alpha^{surv}\cdot T_c^{\max}} > 0$$

Together with the strictly decreasing shape of the survival probability term structure guaranteed by [18], this eliminates any possibility of inconsistency of default and survival probabilities in the exponential spline approximation:

It is worth noting that in most cases the inequality constraints [18], [19] will not be binding and therefore the regression estimates will coincide with the simple GLS formulas. The constraints will kick in precisely in those cases where the input data is not consistent with survival-based modeling, which can happen for variety of reasons including the imperfection of market pricing data, company-specific deviations of expected recovery rates, etc.

We use a two-tiered weighting scheme for the regression objective function with the first set of weights inversely proportional to the square of the bond's spread duration to make sure that the relative accuracy of the hazard rate estimates is roughly constant across maturities. The second set of weights is iteratively adjusted to reduce the influence of the outliers following the generalized M-estimator technique described in Wilcox (1997).

[20]
$$OF = \sum_{q=1}^{\#bonds} w_q^{spread-duration}\cdot w_q^{outlier}\cdot\varepsilon_q^2$$

Equations [12]-[20] fully specify the estimation procedure for survival probability term structure. It satisfies the main goals that we have defined at the outset – the procedure is robust, like any generalized linear regression, it is consistent with market practices and reflects the behavior of distressed bonds, and is guaranteed to provide positive default probabilities and hazard rates.

Figure 3 demonstrates the results of the estimation procedure for the A-rated Industrials, performed monthly for 10 years from July 1994 until June 2004, using the end-of-month prices of senior unsecured bonds in the Lehman Brothers credit database. We show the time series of the 5-year annualized default probability versus the weighted average pricing error of the cross-sectional regression. The latter is defined as the square root of the objective function given by the equation [20], with weights normalized to sum up to 1.

The regression quality has tracked the level of the implied default rates – the higher implied default rates are associated with greater levels of idiosyncratic errors in the cross-sectional regression. This pattern is consistent with the assessment of the issuer-specific excess return volatility during the same period given by the Lehman Brothers multi-factor risk model (see Naldi, Chu and Wang [2002]). We show for comparison the exponentially-weighted specific risk estimates for the A-rated Basic Industries bucket.

**Figure 3. 5-year Implied Default Probability, Average Pricing Error of the Regression, and Risk Model Specific Risk, A-rated Industrials**

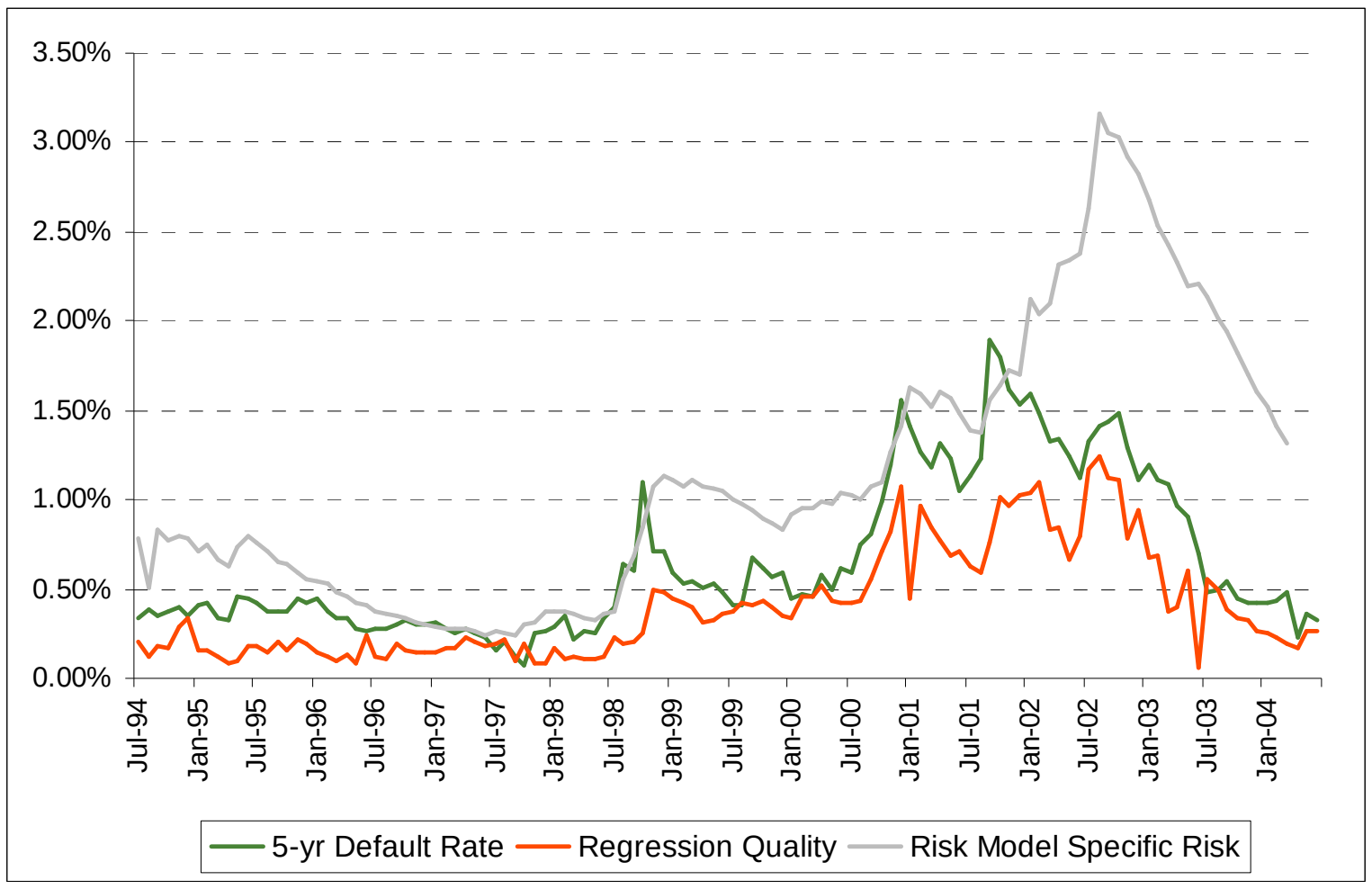

As a final remark we would like to note that the choice of recovery rates used in our model is obviously very important. After all, the main impetus for this methodology was the recognition that recovery rates are a crucial determinant of the market behavior for distressed bonds. In fact, the recovery rate can be considered as a free parameter of the regression and can be estimated from distressed bond prices to yield "market implied" recovery rates. However, the prices of most investment grade bonds do not seem to reflect this information sufficiently to allow for accurate estimation.

Both the cross-sectional, i.e. industry and issuer dependence, and the time-series, i.e. business cycle dependence of the recovery rates is very significant (see Gupton and Stein [2002] and Altman *et. al.* [2002]). Nevertheless, it is often sufficient to use an average recovery rate, such as 40% which is close to the long-term historical average across all issuers, for the methodology to remain robust across the entire range of credit qualities.

## ISSUER AND SECTOR CREDIT TERM STRUCTURES

Having estimated the term structure of survival probabilities, we can now define a set of valuation, risk and relative value measures applicable to collections of bonds such as those belonging to a particular issuer or sector, as well as to individual securities. In practice, to preserve the consistency with market observed bond prices encoded in the survival probability, we define the issuer and sector credit term structures for the same set of bonds which were used in the exponential spline estimation procedures.

As discussed earlier, the conventional Z-spreads are not consistent with survival-based valuation of credit risky bonds. The same can be said about the yield spreads, I-spreads, asset swap spreads, durations, convexities and most other measures which investors currently use day to day. Ultimately, this inconsistency is the source of the breakdown of the conventional spread measures in distressed situations. The market participants know this very well and they stop using these measures for quoting or trading distressed bonds. This situation is commonly referred as "bonds trading on price".

In this Section we will define a host of measures which are consistent with the survival-based approach. We note, however, that the definitions presented in this section do not depend on the specific choice of the exponential splines methodology for fitting survival probability term structures. They can be used in conjunction with any term structure of survival probabilities which is consistent with reduced-form pricing methodology assuming fractional recovery of par – for example one calibrated to the CDS market.

### Implied hazard rate term structure as a valuation measure

Credit investors and market practitioners have long used definitions of spread which correspond to the spread-discount-function methodology, outlined in Section 2. Most commonly used measures, z-spread and OAS, explicitly follow the discounting function approach. Others, such as yield spread or I-spread, implicitly depend on bond-equivalent yields which in turn follow from discounting function approach (see O'Kane and Sen [2004] for definitions). Thus, all of these measures neglect the dependence of the bond price on the recovery value and the debt acceleration in case of default. Therefore, these measures become inadequate for distressed bonds.

In the survival-based approach, spreads are not a primary observed quantity. Only the prices of credit bonds have an unambiguous meaning. Spreads, however we define them, must be derived from the term structure of survival probabilities. There is only one spread measure which is defined directly in terms of survival probabilities. It corresponds to the spread of a hypothetical credit instrument which pays $1 at a given maturity, pays no interest and pays nothing in case of default. We denote this zero-recovery zero-coupon spread as ZZ-spread.

We begin by defining the continuously compounded zero-recovery zero-coupon ZZ-yield as the value which determines the discount price at which such a bond should trade:

[21] $$e^{-Y_{ZZ} \cdot T} = Q(T) \cdot Z_{base}(T), \text{ from where } \quad Y_{ZZ} = -\frac{1}{T} \ln\left(Q(T) \cdot Z_{base}(T)\right)$$

Correspondingly, the risk-free zero-coupon rate is defined as:

[22] $$e^{-y \cdot T} = Z_{base}(T), \quad \text{from where} \quad y = -\frac{1}{T} \ln\left(Z_{base}(T)\right)$$

From these two definitions, we obtain the continuously compounded ZZ-spread as the difference between the ZZ-yield and risk-free zero-coupon rate:

[23] $$S_{ZZ} = Y_{ZZ} - y = -\frac{1}{T}\ln(Q(T))$$

Substituting the definition of the survival probability, we see that the ZZ-spread is equal to the average hazard rate for the maturity horizon of the hypothetical zero-recovery zero-coupon credit bond.

[24] $$S_{ZZ}(T) = \frac{1}{T}\int_0^T h(s)\cdot ds$$

Equivalently, we can say that the instantaneous forward ZZ-spread is identically equal to the hazard rate of the issuer. The hazard rate also has a meaning of instantaneous forward default probability, i.e. the probability intensity of default during a small time interval in the future provided that the issuer has survived until that time. Thus, the forward ZZ-spread is equal to the forward default probability:

[25] $$S_{ZZ}^{fwd}(t) = h(t)$$

Note that this is the same relationship as the one used in the standard implementations of the reduced-form models of default (Litterman and Iben [1991], Jarrow and Turnbull [1995], Duffie and Singleton [1997]). The novelty of our approach is that we do not estimate the hazard rate from spread curves, but conversely, derive the spread curves from hazard rates that are obtained directly from the bond prices via the fitted survival probability. Since the conventional spread curve estimates are not consistent with the survival-based approach, deriving hazard rates from such spread measures is fraught with inaccuracies and biases, especially for high hazard rates. Only if the recovery rate is equal to zero and for the case of zero coupon credit bonds the conventional Z-spread becomes equal to the forward ZZ-spread and becomes consistent with the survival-based valuation – in concordance with our earlier assertions.

Using the exponential spline representation of the survival probability term structure [12], we can derive the hazard rate term structure as follows:

[26] $$h(t) = -\frac{d}{dt}\ln Q(t) = \frac{\sum_{k=1}^{3} k\cdot\alpha\cdot\beta_k^{surv}\cdot e^{-k\cdot\alpha^{surv}\cdot t}}{\sum_{k=1}^{3}\beta_k^{surv}\cdot e^{-k\cdot\alpha^{surv}\cdot t}}$$

This formula confirms again that the inequality constraint [18] indeed guarantees the positivity of the hazard rates.

Figure 4 shows the result of estimation of the hazard rate (forward ZZ-spread) term structures for Ford and for the BBB Consumer Cyclicals sector.

**Figure 4. Hazard rates (forward ZZ-spreads) for Ford and BBB Consumer Cyclicals (as of 12/31/03)**

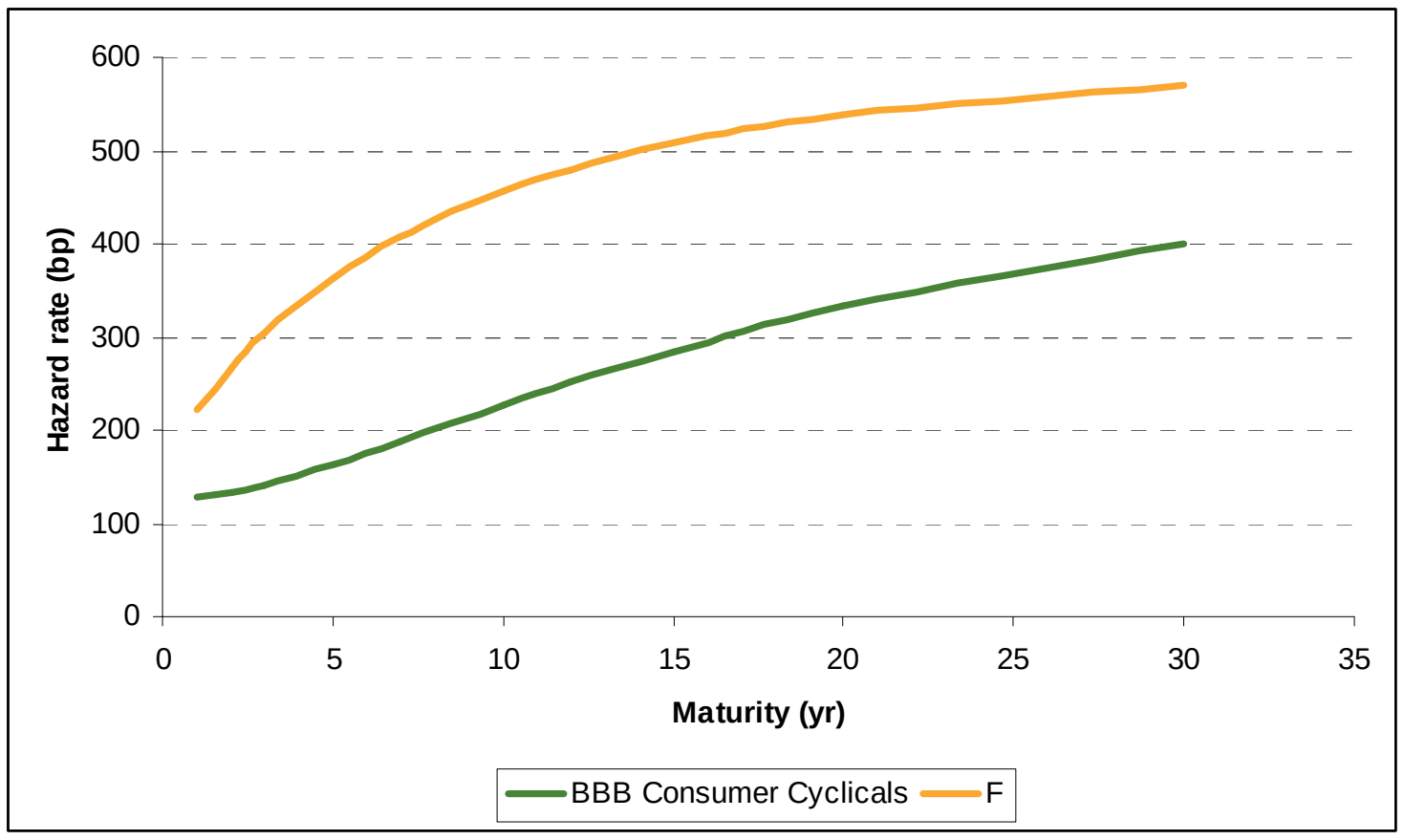

Even though we classify the hazard rate primarily as a valuation measure, it can be used for relative value assessments. It is important to remember that the implied hazard rate does not correspond to an actual forecast. Like other market-implied parameters, it also incorporates in a complex way a host of risk premia which reflect both credit and non-credit factors such as recovery rate risk, liquidity, etc. Nevertheless, fitted hazard rate term structures may provide valuable clues beyond the conventional spread analysis. For example, the Ford curve in Figure 4 is substantially wider than the sector curve, signalling the higher credit risk associated with Ford bonds. The Ford curve has a distinctive shape, with the maximum differential to sector curve in intermediate maturities, suggesting that these maturities offer the best opportunities for monetizing views on the relative risk and return between this issuer and the industry sector.

## Par coupon and P-spread term structures

While the definition of the ZZ-spread presented in the previous subsection is quantitatively sound, it is not likely to be of much value for practitioners because the zero-coupon zero-recovery credit bonds do not actually exist in the marketplace. While there are few pure discount (zero-coupon) securities, particularly in the short term CP market, all of them are subject to equal-priority recovery rules and therefore cannot be considered zero-recovery. On the other hand, there exist credit derivatives such as digital default swaps which can have a contractual zero recovery, but they do have premium payments and therefore cannot be considered as an equivalent of a zero-coupon bond. Therefore, the usefulness of the ZZ-spread as a relative value measure is limited – one can obtain from it some insight about the issuer or the sector but not about a particular security.

The vast majority of the credit market consists of interest-bearing instruments subject to equal-priority recovery in case of default. A practically useful relative value measure should refer to such instruments and should contrast them with credit risk-free instruments such as Treasury bonds or interest rate swaps which provide a funding rate benchmark[4]. One such measure is the par coupon – i.e. a coupon of a hypothetical bond of a given maturity which would trade at a par price if evaluated using the fitted issuer survival probability term structure. Correspondingly, the par spread is defined by subtracting the fitted par yield of risk-free bonds from the fitted par coupon of credit-risky bonds.

*[4] Due to collateral rules and strict margin requirements for interest rate swap trading, the swaps can be considered a credit risk-free instrument. While this statement is not true regarding a specific over-the-counter trade which is subject to counterparty risk, the market-wide average swap rates bear little of issuer-specific credit risk. They may, however, be exposed to systemic credit risks affecting the financial sector.*

One must note that the par coupon and par spread measures do not correspond to a specific bond – these are derived measures based on the issuer survival curve and a specific price target equal to par, i.e. 100% of face value. The par price of the bond has a special significance, because for this price the expected price return of a risk-free bond to maturity is precisely zero. Therefore, the par yield of a risk-free bond reflects its expected total return to maturity and the par spread of a credit-risky bond reflects its (risky) excess return to maturity. Thus, the par spread defined above can be considered a consistent (fair) relative value measure for a given issuer/sector for a given maturity horizon.

For example, if two different issuers wanted to buy back their outstanding bonds in the secondary market and instead issue new par bonds of the same maturity $T$ then the fair level of the coupons which the market should settle at, assuming no material change in the issuers' credit quality, would be given by the respective fitted par coupons. Correspondingly, investors considering these new bonds can expect excess returns to maturity $T$ equal to their respective fitted par spreads. If one of those spreads is greater than the other – this would represent a relative value which the investors should contrast with their views of the issuers' credit risks to the said maturity horizon.

For bonds with coupon frequency $f$ (usually annual, $f=1$, or semi-annual, $f=2$) with an integer number of payment periods until maturity $t_N = N \cdot \frac{1}{f}$ we define the par coupon term structure by solving for the coupon level from the pricing equation [11].

$$[27] \quad C^{par}(t_N | f) = f \cdot \frac{1 - Q(t_N) \cdot Z_{base}(t_N) - R_p \cdot \sum_{i=1}^{N} \left(Q(t_{i-1}) - Q(t_i)\right) \cdot Z_{base}(t_i)}{\sum_{i=1}^{N} Q(t_i) \cdot Z_{base}(t_i)}$$

Contrast this definition with what the one that would be consistent with spread discount function based approaches:

$$[28] \quad C_{SDF}^{par}(t_N | f) = f \cdot \frac{1 - Z_{base}(t_N) \cdot Z_{spread}(t_N)}{\sum_{i=1}^{N} Z_{base}(t_i) \cdot Z_{spread}(t_i)}$$

We can see that the latter definition only coincides with the former if we assume that the recovery rate is zero and that the spread discount function is equal to the survival probability.

An example of the fitted par coupon term structure is shown in Figure 5 where we show our estimates for Ford and BBB Consumer Cyclicals sector as of 12/31/2003. Note that the shape of these fitted curves – very steep front end and flattish long maturities – is largely determined by the shape of the underlying risk-free curve (we used the swaps curve in this case), with the credit risk being a second-order modification for most issuers and sectors, except those that trade at very deep discounts.

**Figure 5. Fitted par coupon for Ford and BBB Consumer Cyclicals (as of 12/31/03)**

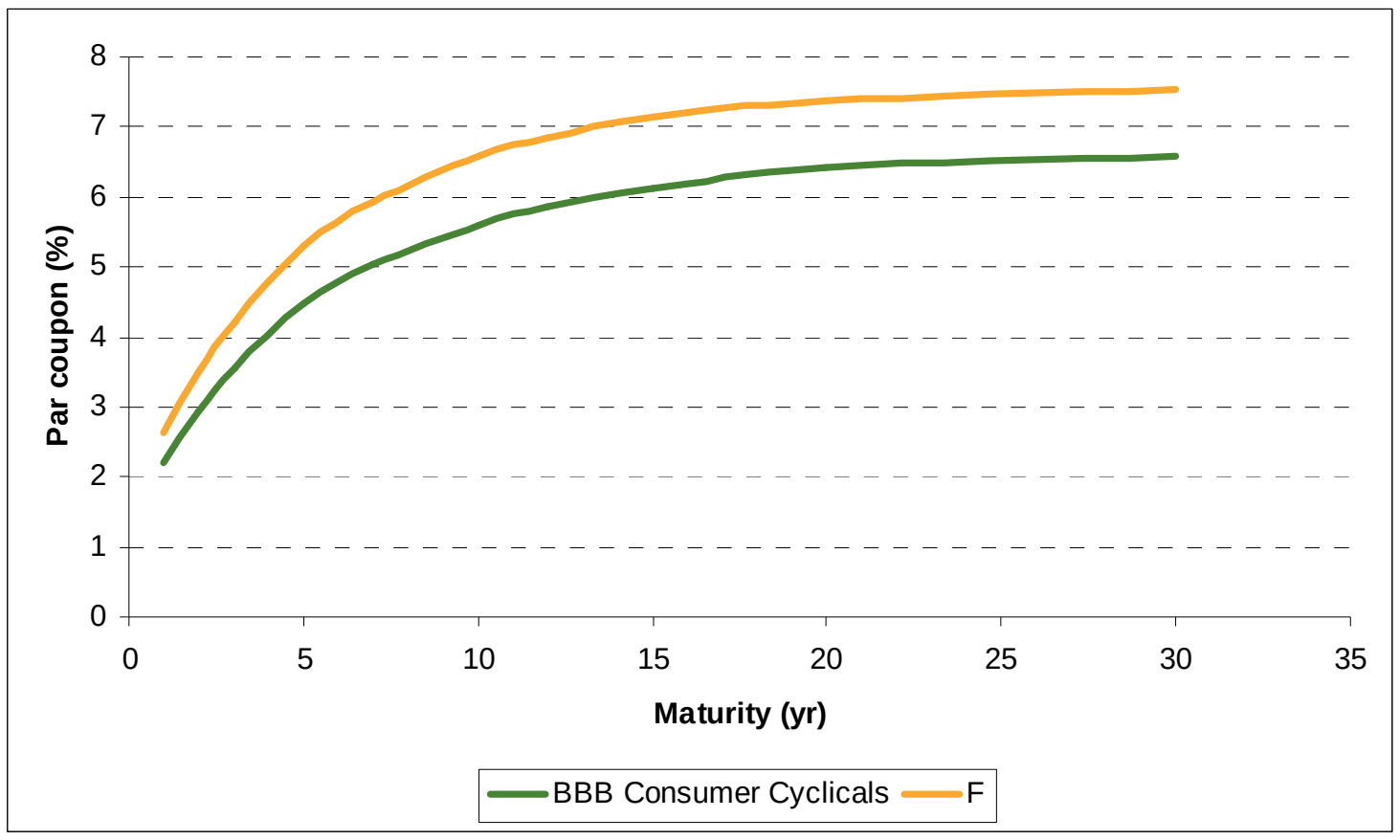

Let us also define the par yield of the risk-free bond in a similar fashion:

[29] $$Y_{base}^{par}(t_N | f) = f \cdot \frac{1 - Z_{base}(t_N)}{\sum_{i=1}^{N} Z_{base}(t_i)}$$

The par spread (P-spread) to the base curve (either Treasury or swaps) can then be derived by subtracting the par base yields from the par risky coupons of the same maturities:

[30] $$S_{base}^{par}(T | f) = C^{par}(T | f) - Y_{base}^{par}(T | f)$$

Figure 6 demonstrates the fitted par Libor spread term structures using the same example of the Ford and BBB Consumer Cyclicals sectors.

**Figure 6. Fitted Libor P-spread for Ford and BBB Consumer Cyclicals (as of 12/31/03)**

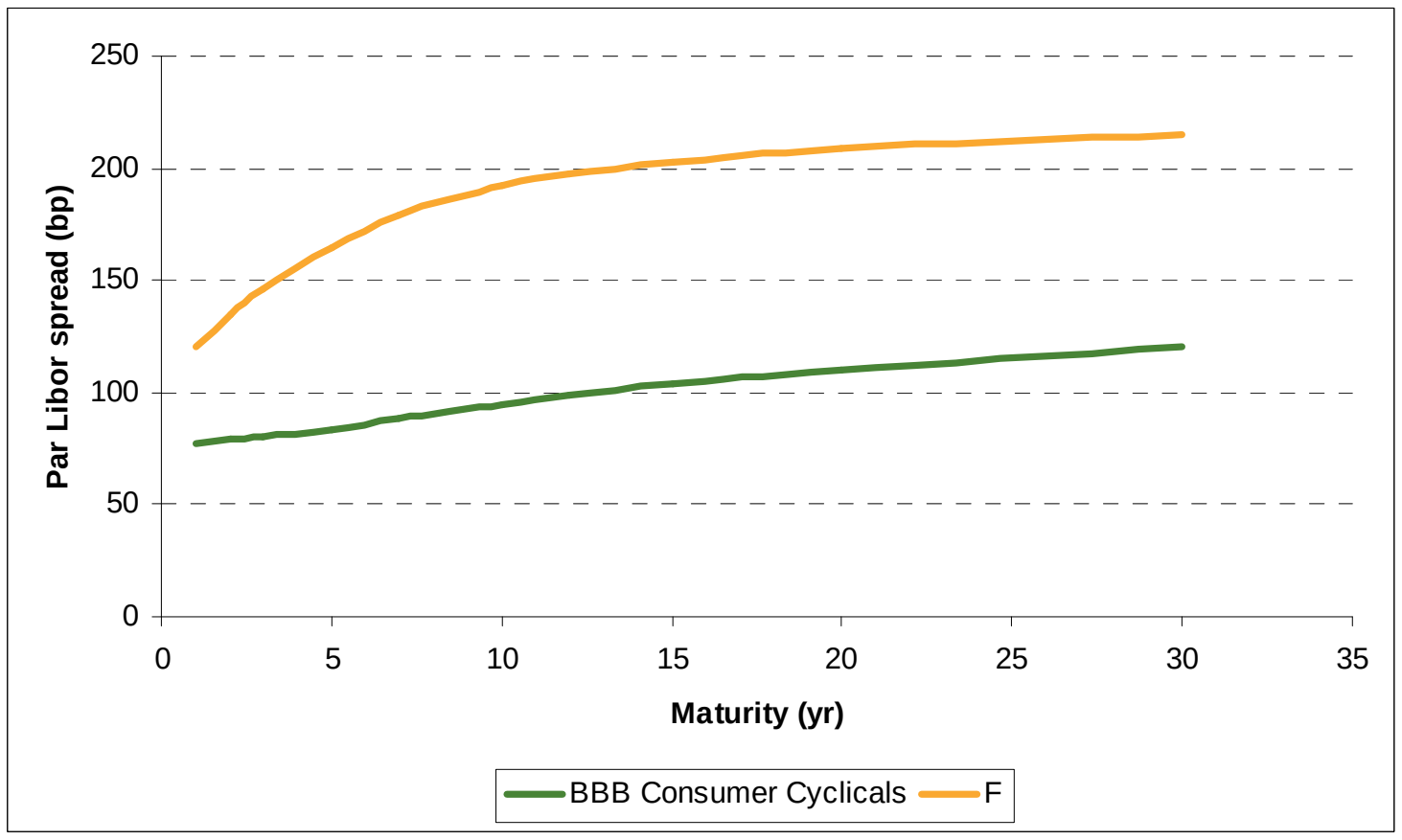

### Bond-Implied CDS (BCDS) spread term structure

The survival-based valuation approach is well suited for the CDS market. In fact it has been the market practice since its inception. By deriving the bond-implied CDS spreads within the same framework we are aiming to give investors an apples-to-apples relative value measure across the bond and CDS markets.

The pricing relationship for credit default swaps simply states that the expected present value of the premium leg is equal to the expected present value of the contingent payment.

[31] $$\frac{S}{f}\cdot\sum_{i=1}^{N} Q(t,t_i)\cdot Z(t,t_i) = (1-R)\cdot\sum_{i=1}^{N}\left(Q(t,t_{i-1})-Q(t,t_i)\right)\cdot Z(t,t_i)$$

where, $S$ is the annual spread, $f$ is the payment frequency ( $f=4$ ), $Q(t,t_i)$ is the issuer's survival probability between time $t$ to time $t_i$, $Z(t,t_i)$ stands for the (default risk-free) discount factor from time $t$ to time $t_i$, and $R$ stands for the recovery rate. We have used for simplicity a discrete-time approximation which coincides with the quarterly coupon payment dates (see O'Kane and Turnbull [2003] for an exact treatment of CDS).

The bond-implied CDS spread term structure, hereafter denoted as BCDS term structure, is defined by substituting the survival probability term structure fitted from bond prices, $Q_{bond}(t,t_i)$ , into the following equation for par CDS spreads:

[32] $$BCDS(t,t_N) = f\cdot(1-R)\cdot\frac{\sum_{i=1}^{N}\left(Q_{bond}(t,t_{i-1})-Q_{bond}(t,t_i)\right)\cdot Z(t,t_i)}{\sum_{i=1}^{N} Q_{bond}(t,t_i)\cdot Z(t,t_i)}$$

The BCDS term structure gives yet another par-equivalent measure of spread for credit-risky issuers in addition to the par spread. We will show in Part 3 of this series that there exists a complementarity between the BCDS term structure and the properly defined credit-risk-free benchmark security, proving that the BCDS spread is a clean measure of excess return, free of biases associated with non-par prices.

### Constant coupon price (CCP) term structures

Since we argued that the price-based estimation techniques are more consistent than those fitting yields or spreads, it is useful to define a set of credit term structures expressed in terms of bond prices. For any integer number of payment periods $t_N = N\cdot\frac{1}{f}$, we define the constant coupon price (CCP) term structure as the price level of the bond with a pre-set coupon (e.g., coupons of 6%, 8%, 10%).

[33] $$\begin{aligned} P(t_N|C) \;=\; & \Bigg[\sum_{i=1}^{N}\frac{C}{f}\cdot Z_{base}(t_i)\cdot Q(t_i) + Z_{base}(t_N)\cdot Q(t_N) \\ & + \sum_{i=1}^{N} R_p\cdot Z_{base}(t_i)\cdot\left(Q(t_{i-1})-Q(t_i)\right)\Bigg] \end{aligned}$$

Figure 7 shows estimated CCP term structures for Georgia Pacific as of December 31, 2002. The prices are calculated as fractions of a 100 face, i.e. par price 100% appears as 100. We observe that the 6%, 8% and 10% Constant Coupon Price term structures neatly envelope the scatterplot of prices of individual bonds which have coupon levels ranging from 6.625% to 9.625%. Importantly, all three CCP term structures correspond to *the same* term structure of survival probability, and thus embody the same credit relative value. To the extent that a price of a bond with a particular fixed coupon is in line with the level suggested by the CCP term structures, this bond also reflects the same credit relative value. Thus, a graph like this can serve as a first crude indication of the relative value across bonds of a given issuer or sector – especially when there are securities whose prices are substantially different from the corresponding fitted price curve levels. In the next section we develop a more precise measure for assessing such relative value, which we call OAS-to-Fit.

**Figure 7. CCP term structures and bond prices, Georgia Pacific as of 12/31/02**

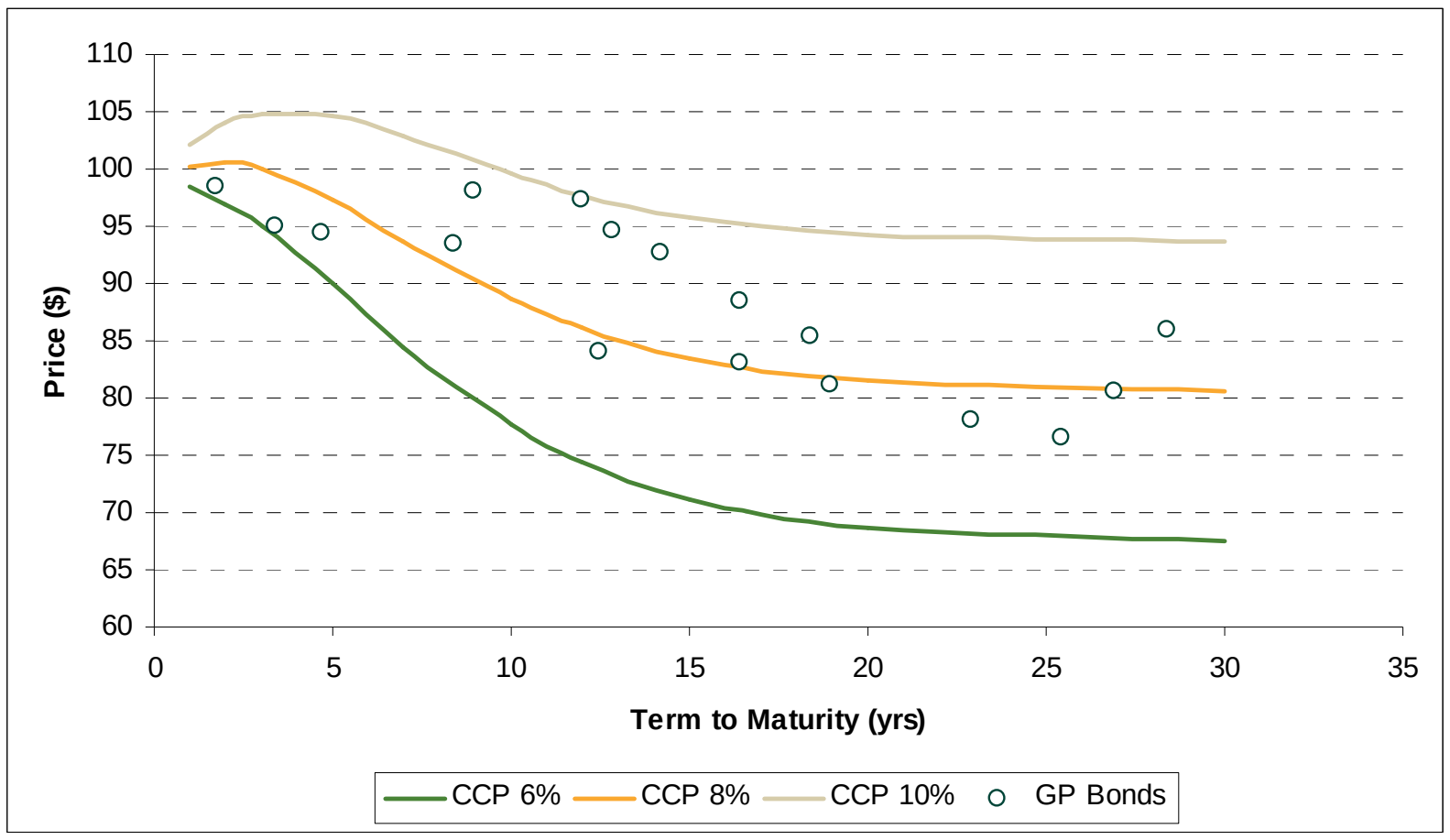

Note also how the CCP term structures tend to become flat at longer maturities – this is a reflection of the fact that Georgia Pacific was trading at elevated levels of the implied default risk, with the fitted hazard rate exceeding 1000 bp at maturities longer than 5 years. At such levels of credit risk the implied survival probability to 10 years or longer is only about 35%, and the recovery scenario at longer maturities becomes the dominant one. This, in turn, leads to a flat term structure of prices.

One might say that the reason for this is that the high probability of an early default scenario causes the bonds to trade with an effective life much shorter than their nominal maturity. We show in the next section that the properly defined duration measure for credit bonds that is consistent with the survival-based valuation will exhibit the same feature, with the duration becoming much shorter for bonds with higher levels of the implied credit risk.

## BOND-SPECIFIC MEASURES

So far we have developed a set of term structures which encode our knowledge about the issuer (or sector) as a whole, rather than about the specific bond. In particular, our primary measure, the term structure of survival probability, clearly refers to the issuer and not to any particular bond issued by this issuer. It would make no sense to say that the XYZ 6.5% bond maturing in 10 years has a term structure of survival probabilities, but it does make sense to say that the XYZ issuer has a term structure of survival probabilities which was fitted using the price of the above mentioned bond, as well as other bonds of the same issuer, if available.

When it comes to a particular bond, investors are typically concerned with their “fair value” and relative value with respect to other bonds of the same issuer or sector. The estimate of the fair value for a given bond is a straightforward application of the issuer- or sector-specific fair value metrics to the particular maturity and coupon of the security under investigation. The answer to the second question lies in the comparison of the market-observed bond price with the estimated fair value price. The OAS-to-Fit (OASF) measure, introduced below, provides an unambiguous and consistent way to make such a comparison, free of biases associated with the term to maturity or level of coupon, which plague the conventional spread measures.

## Bond's fitted price and fitted par coupon

The CCP measure introduced in the previous section determines the precise fit of the bond's price for a hypothetical bond with a pre-set coupon and maturity chosen so that there are integer number of payment periods and no accrued coupon amount as of pricing date.

We can easily extend this generic fitted price measure to any given bond by defining it as the clean price such a bond would have if it was priced precisely by the issuer- (or sector) fitted survival probability term structure. Clearly, the notion of the "fitted" price depends on the context of the "fit", i.e. whether we are talking about the fair value with respect to issuer or sector. A bond can be undervalued with respect to other bonds of the same issuer, but overvalued with respect to majority of the bonds in the larger industry or rating sector.

Since the accrued interest is a known value depending on the coupon level and pricing date, the fitted price in our implementation is precisely equal to the market price less the regression residual. In other implementations one would calculate it using the term structure of the survival probability which is considered to reflect the "fair" value for the issuer, such as the one calibrated from the benchmark CDS spread levels:

[34]
$$\begin{aligned} P_{fit} &= \sum_{i=1}^{N} \frac{C}{f} \cdot Z_{base}(t_i) \cdot Q(t_i) + Z_{base}(t_N) \cdot Q(t_N) \\ &+ \sum_{i=1}^{N} R_p \cdot Z_{base}(t_i) \cdot \left( Q(t_{i-1}) - Q(t_i) \right) \\ &- A_{\text{int}} \end{aligned}$$

The fitted par coupon of a given bond is defined as the coupon which would make this bond's clean price equal to par when evaluated using the suitably chosen fitted survival term structure. For a given maturity date, we must modify equation [31] for the par coupon to account for the effect of the non-current coupon and the non-zero accrued interest amount $A(C) = C \cdot T_{accrued}$:

[35]
$$C_{fit}^{par} = \frac{1 - Q(t_N) \cdot Z_{base}(t_N) - R_p \cdot \sum_{i=1}^{N} \left( Q(t_{i-1}) - Q(t_i) \right) \cdot Z_{base}(t_i)}{\left( \frac{1}{f} \sum_{i=1}^{N} Q(t_i) \cdot Z_{base}(t_i) \right) - T_{accrued}}$$

The difference between the fitted par coupon and the correspondingly defined fitted par base (LIBOR or Treasury) rate for the same maturity could be termed the "fair P-spread".

[36]
$$S_{fit}^{par}(T) = C_{fit}^{par}(T) - Y_{base}^{par}(T)$$

This spread, however, does not correspond to the observed price of the bond. It is instead a P-spread that the bond would have if it was priced precisely, without any residual errors, by the corresponding issuer- or sector-specific fitted survival probability term structure. Next, we turn our attention to pricing errors and relative value measures.

## OAS-to-Fit (OASF) as a relative value measure

For a long time credit investors have been using spread measures such as nominal spread, I-spread, OAS or Z-spread, to assess the relative value across various bonds of the same issuer or sector (see O'Kane and Sen [2004] for a glossary of terms). In the previous sections we have demonstrated that these measures have inherent biases because all of them rely on the strippable cash flow valuation assumption which is inadequate for credit-risky bonds.

We argued that spread-like measures which can be interpreted in terms of risky excess return to maturity and are in agreement with the survival-based valuation, correspond to idealized par bonds or par-equivalent instruments such as CDS. However, bonds in the secondary market typically trade away from par. It is important to measure the degree by which the bond's price deviates from the "fair" price corresponding to its coupon level and the term structure of the underlying interest rates. As explained in the previous subsection, the latter is the analogue of the constant coupon price term structure. For example, in terms of credit relative value a 6% bond trading at a price close to the 6% CCP curve is not any different from a 10% bond trading close to 10% CCP curve. On the other hand, if the first bond were trading above the 6% CCP price while the second was trading below the 10% CCP price, we would say that the first bond is rich and the second bond is cheap, even though the observed price of the first bond might be less than that of the second bond.

Fortunately, our estimation methodology for survival probability term structures lends itself naturally to a robust determination of the rich/cheap measures as described above. Indeed, we impose a fairly rigid structural constraint of the shape of the survival probability term structure by adopting the exponential splines approximation. The result was that the cross-sectional regression which gives the spline coefficients does not in general price any of the bonds precisely. Instead, we make a trade-off between the individual bond pricing precision and the robustness of the overall fit. The bond-specific pricing errors from the best fit of the survival probability term structure (i.e. cross-sectional regression residuals) then become a natural candidate for the relative value across the bonds – if the residual is positive we say that the bond is rich, and if the residual is negative we say that it is cheap.

We can express the pricing error in terms of a constant OAS-to-Fit which acts as an additional discount (or premium) spread that replicates the bond's present value:

[37]
$$\begin{aligned} P_{mkt} + A_{\text{int}} \quad = \quad & \sum_{i=1}^{N} \frac{C}{f} \cdot Z_{base}(t_i) \cdot Q(t_i) \cdot e^{-OASF \cdot t_i} + Z_{base}(t_N) \cdot Q(t_N) \cdot e^{-OASF \cdot t_N} \\ + \quad & \sum_{i=1}^{N} R_p \cdot Z_{base}(t_i) \cdot \left( Q(t_{i-1}) - Q(t_i) \right) \cdot e^{-OASF \cdot t_i} \end{aligned}$$

Here, $P_{mkt}$ is the observed market price, and $A_{\text{int}}$ is the accrued interest. OAS-to-Fit should be interpreted as a bond-specific premium/discount which reflects both market inefficiencies such as liquidity premia and biases related to persistent market mispricing of credit bonds. We say that a positive OASF (negative regression residual) signals cheapness and the negative OASF (positive residual) signals richness across the bonds of the same issuer. This is similar to the usual meaning assigned to spreads. Moreover, we can say that the OASF differential for two bonds reflects the "clean" relative value between them.

Correspondingly, the market P-spread of the bond is equal to the previously derived "fair P-spread" plus the OAS-to-Fit (to recover the observed pricing of the bond). The par spread can be defined with respect to any base credit risk-free curve. We write down below the definition corresponding to the LIBOR spread:

[38]
$$S_{fit}^{bond}(T) = C_{fit}^{par}(T) - Y_{base}^{par}(T) + OASF^{bond}$$

Figure 8 shows the OASF for Calpine bonds as of 6/30/2003. OASF ranges from -97 bp to +107bp, indicating a potential for as much as 200 bp relative value between the bonds, even for a set of bonds which have fairly close maturities and similar liquidity.

**Figure 8. OAS-to-Fit for Calpine bonds as of 6/30/03**

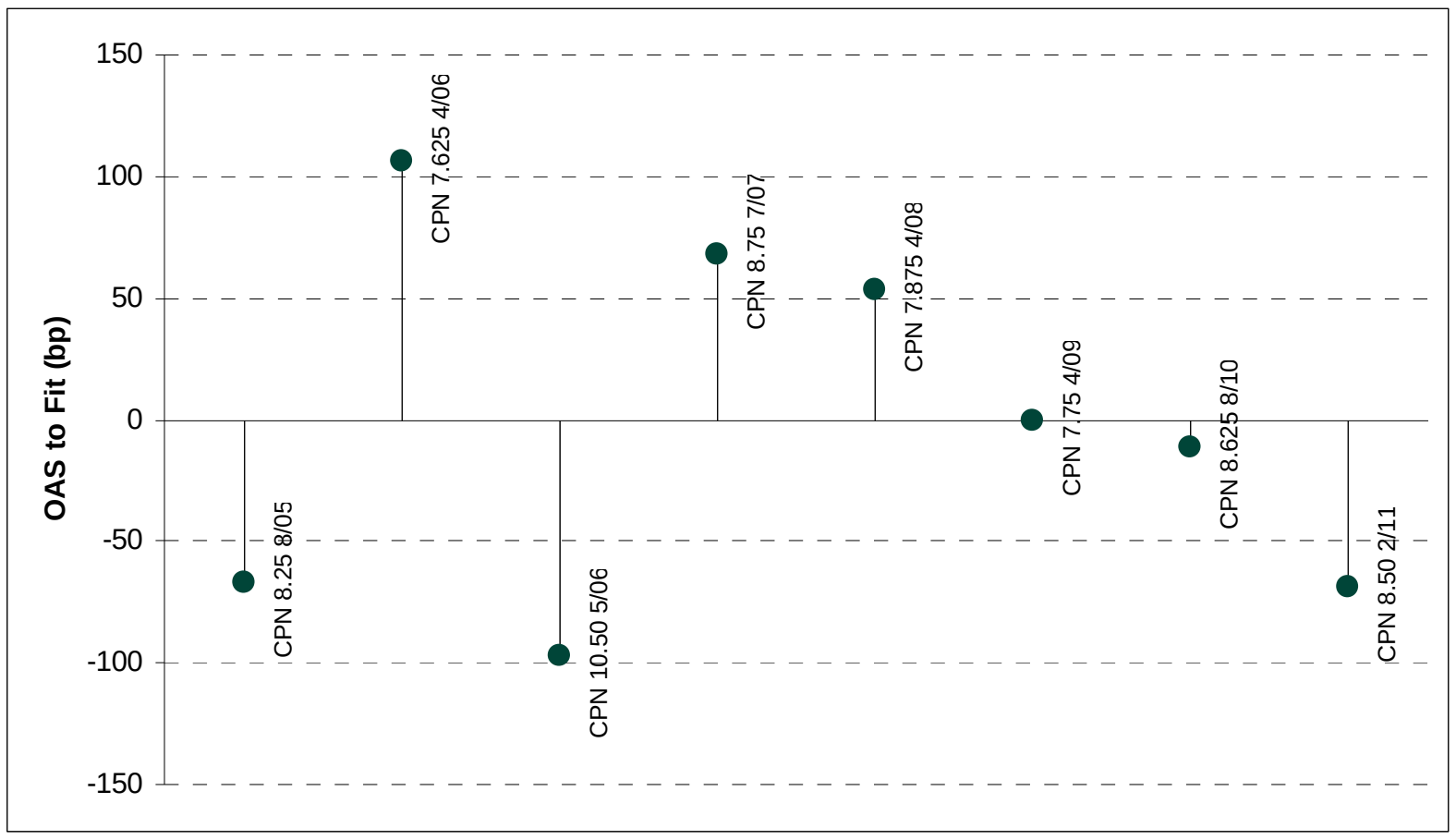

Figure 9 tabulates the calculated values for the same Calpine bonds as shown in Figure 8. Note the large differences between the Z-spread and P-spread measures of the bonds across all maturities, ranging from 250 to 400 bp. Note also that looking at Z-spreads alone it would be difficult to distinguish between the fair valuation following a spread term structure and bond-specific relative value. By contrast, OASF provides a clean measure of such relative value. It appears that most high-coupon bonds are rich (have negative OASF), while the low-coupon bonds are cheap (have positive OASF). Such patterns of relative value driven by the coupon levels are quite often observed, and are a consequence of the market mispricing driven by the use of nominal spreads and conventional OAS.

**Figure 9. Calpine bonds as of 6/30/03**

| *Description* | *Maturity (yrs)* | *Coupon* | *Z-Spread* | *P-Spread* | *OASF* | *Price* | *Fitted Price* | *Price Residual* |
|---|---|---|---|---|---|---|---|---|
| CPN 8.25 8/05 | 2.13 | 8.25 | 1649 | 1949 | -68 | 82.00 | 81.02 | 0.98 |
| CPN 7.625 4/06 | 2.79 | 7.63 | 1690 | 2061 | 107 | 75.00 | 76.75 | -1.75 |
| CPN 10.50 5/06 | 2.82 | 10.50 | 1594 | 1813 | -97 | 83.30 | 81.59 | 1.71 |
| CPN 8.75 7/07 | 3.77 | 8.75 | 1568 | 1807 | 68 | 74.52 | 75.86 | -1.34 |
| CPN 7.875 4/08 | 4.76 | 7.88 | 1376 | 1781 | 55 | 71.00 | 72.18 | -1.18 |
| CPN 7.75 4/09 | 5.79 | 7.75 | 1210 | 1610 | 0 | 71.00 | 71.00 | 0.00 |
| CPN 8.625 8/10 | 7.13 | 8.63 | 1110 | 1457 | -13 | 73.50 | 73.15 | 0.35 |
| CPN 8.50 2/11 | 7.63 | 8.50 | 1023 | 1353 | -71 | 75.00 | 72.93 | 2.07 |

## CONCLUSIONS

In this paper we have critically examined the conventional bond pricing methodology and have shown that it does not adequately reflect the nature of the credit risk faced by investors. In particular, we have demonstrated that the strippable discounted cash flows valuation assumption which is normally taken for granted by most analysts, leads to biased estimates of relative value for credit bonds. Moreover, even the CDS market which does not use this pricing methodology is strongly influenced by its prevalence in the cash market because of the constant cross-benchmarking of cash bonds and CDS. The example of the “optically distorted” term structures of Z-spreads and CDS spreads of such a highly liquid name as Ford Motor Credit shown in section 2 should suffice to convince investors that the conventional methodology can indeed be quite misleading.

We have introduced a consistent survival-based valuation methodology which is free of the biases mentioned above, albeit at a price of abandoning of the strippable discounted cash flows valuation assumption. We also developed a robust estimation methodology for survival probability term structures using the exponential splines approximation, and implemented and tested this methodology in a wide variety of market conditions and across a large set of sectors and issuers, from the highest credit quality to highly distressed ones.

To remedy the loss of intuition due to the abandonment of OAS, Z-spreads, and other conventional spread measures, we have introduced a host of new definitions for credit term structures, ranging from valuation measures such as the hazard rate and P-spread to relative value measures such as the bond-specific OAS-to-Fit.

In conclusion, we believe that the eventual adoption of the survival-based methodologies advocated in this paper by market participants will lead to an increase in the efficiency of the credit markets just as the adoption of better prepayment models led to efficiency in the MBS markets twenty years ago. Investors who will be at the forefront of this change will be in a position to benefit from the secular shift to a much more quantitative approach to credit portfolio management. We have witnessed many such turning points in recent years, including the widespread following attained by structural credit risk models pioneered by Merton and further developed by KMV and others, the explosive growth in the credit derivatives and structured credit market which proceeded alongside with a dramatic progress in modelling complex risks of correlated defaults and losses, the adoption of new banking regulatory standards which place a much greater emphasis on quantitative measures of credit risk, and finally the proliferation of credit hedge funds and relative value investors who stand ready to exploit any inefficiencies still present in the marketplace. It was long overdue that the most traditional of all credit instruments, the credit bonds, have also been considered in the light of this new knowledge. We hope that our paper contributes to this worthy task.

*Acknowledgements:* We would like to thank Marco Naldi as well as many other colleagues at Lehman Brothers Fixed Income Research department for numerous helpful discussions throughout the development and implementation of the survival-based methodology during the past several years.

## REFERENCES

**Adams, K. J. and D. R. van Deventer (1994):** *Fitting Yield Curves and Forward Rate Curves with Maximum Smoothness*, Journal of Fixed Income, June issue, p. 52

**Altman, E. I., B. Brooks, A. Resti and A. Sironi (2002):** *The Link between Default and Recovery Rates: Theory, Empirical Evidence and Implications*, working paper, NYU Stern School of Business

**Bakshi, G., D. Madan, and F. Zhang (2004):** *Understanding the Role of Recovery in Default Risk Models: Empirical Comparisons and Implied Recovery Rates*, working paper, R. H. Smith School of Business, University of Maryland

**Duffie, D. (1998):** *Defaultable Term Structure Models with Fractional Recovery of Par*, working paper, Graduate School of Business, Stanford University

**Duffie, D., M. Schroder and C. Skiadas (1996):** *Recursive Valuation of Defaultable Securities and the Timing of Resolution of Uncertainty*, Annals of Applied Probability, vol. 6 (4), p. 1075

**Duffie, D. and K. Singleton (1999):** *Modeling Term Structures of Defaultable Bonds*, Review of Financial Studies, vol. 12, p. 687

**Duffie, D. and K. Singleton (2003):** *Credit Risk*, Princeton University Press

**Finkelstein, V. (1999):** *The Price of Credit*, Risk, vol. 12 (12), p. 68

**Guha, R. (2002):** *Recovery of Face Value: Theory and Empirical Evidence*, working paper, London Business School

**Gupton, G. and R. M. Stein (2002):** *LossCalc: Moody's Model for Predicting Loss Given Default*, Moody's Special Comment, February 2002

**Jarrow, R. A. (2001):** *Default Parameter Estimation Using Market Prices*, Financial Analyst Journal, vol. 57 (5), p. 75

**Jarrow, R. A. and S. M. Turnbull (1995):** *Pricing Options on Financial Securities Subject to Default Risk*, Journal of Finance, vol. 50, p. 53

**Jarrow, R. A. and S. M. Turnbull (2000):** *The Intersection of Market and Credit Risk*, Journal of Banking and Finance, vol. 24, p. 271

**Jarrow, R. A., D. Lando and S. M. Turnbull (1997):** *A Markov Model for the Term Structure of Credit Risk Spreads*, Review of Financial Studies, vol. 10, p. 481

**Jarrow, R. A. and Y. Yildirim (2002):** *Valuing Default Swaps under Market and Credit Risk Correlation*, Journal of Fixed Income, vol. 11 (4), p. 7

**Litterman, R. and T. Iben (1991):** *Corporate Bond Valuation and the Term Structure of Credit Spreads*, Journal of Portfolio Management, vol. 17 (3), p. 52

**Merton, R. (1974):** *On the Pricing of Corporate Debt: the Risk Structure of Interest Rates*, Journal of Finance, vol. 29, p. 449

**Naldi, M., K. Chu and G. Chang (2002):** *The New Lehman Brothers Credit Risk Model*, Quantitative Credit Research Quarterly, vol. 2002-Q2, Lehman Brothers

**O'Kane, D., and S. Sen (2004):** *Credit Spreads Explained*, Quantitative Credit Research Quarterly, vol. 2004-Q1, Lehman Brothers

**O'Kane, D., and S. M. Turnbull (2003):** *Valuation of Credit Default Swaps*, Quantitative Credit Research Quarterly, vol. 2003-Q1/Q2, Lehman Brothers

**Shea, G. S. (1985):** *Term Structure Estimation with Exponential Splines*, Journal of Finance, vol. 40, p. 319

**Schonbucher, P. J. (2003):** *Credit Derivatives Pricing Models*, John Wiley & Sons

**Vasicek, O. and G. Fong (1982):** *Term Structure Modeling Using Exponential Splines*, Journal of Finance, vol. 37, 339

**Wilcox, R. (1997):** *Introduction to Robust Estimation and Hypothesis Testing*, Academic Press

**Zhou F. (2002):** *The Swap Curve*, Lehman Brothers Fixed Income Research

## APPENDIX: DEFINITION OF EXPONENTIAL SPLINES

The exponential spline approximation is defined in a way which facilitates smooth fitting of functions which are exponentially decreasing with term to maturity but are not necessarily required to have a constant rate of decrease. The shape of the approximated function is given by a linear combination of spline factors:

[39] $$F(t) = \sum_{k=1}^{K_{treas}} \beta_k \cdot Spline_k(t|\alpha)$$

where the spline coefficients $\beta_k$ are constants derived from the best fit optimization procedure, such as minimization of the weighted average square of bond pricing errors in the case of fitting Treasury discount functions or credit survival probability term structures.

The exponential spline component functions $Spline_k(t|\alpha)$ have a fixed shape depending only on the remaining term to maturity, and on the decay parameter $\alpha$.

The first three spline factors are known as no-knot factors because they are smooth in the entire range of maturities.

[40] $$Spline_{k\le 3}(t|\alpha) = e^{-k\cdot\alpha\cdot t}$$

In fixed income applications, the shape of the yield curve often reflects market segmentation – the short, medium and long maturities can have substantially different behavior. To address this, one uses higher order spline factors (number 4 and above) which are exactly zero below certain maturity known as the "knot point" $T_k^{knot}$, have the familiar exponential shape above the knot point, and have a smooth value and first derivative at the knot point itself. These requirements determine the higher order spline factors uniquely as follows:

[41] $$Spline_{k\ge 4}(t|\alpha) = \Theta(t - T_k^{knot}) \cdot \left( \frac{1}{3} + e^{-2\cdot\alpha\cdot(t-T_k^{knot})} - e^{-\alpha\cdot(t-T_k^{knot})} - \frac{1}{3} e^{-3\cdot\alpha\cdot(t-T_k^{knot})} \right)$$

The shapes of the spline factors are shown in Figure A1 for no-knot factors (order 1 through 3), and in Figure A2 for the knot factors of order 4 and above.

**Figure A1. No-knot spline factors**

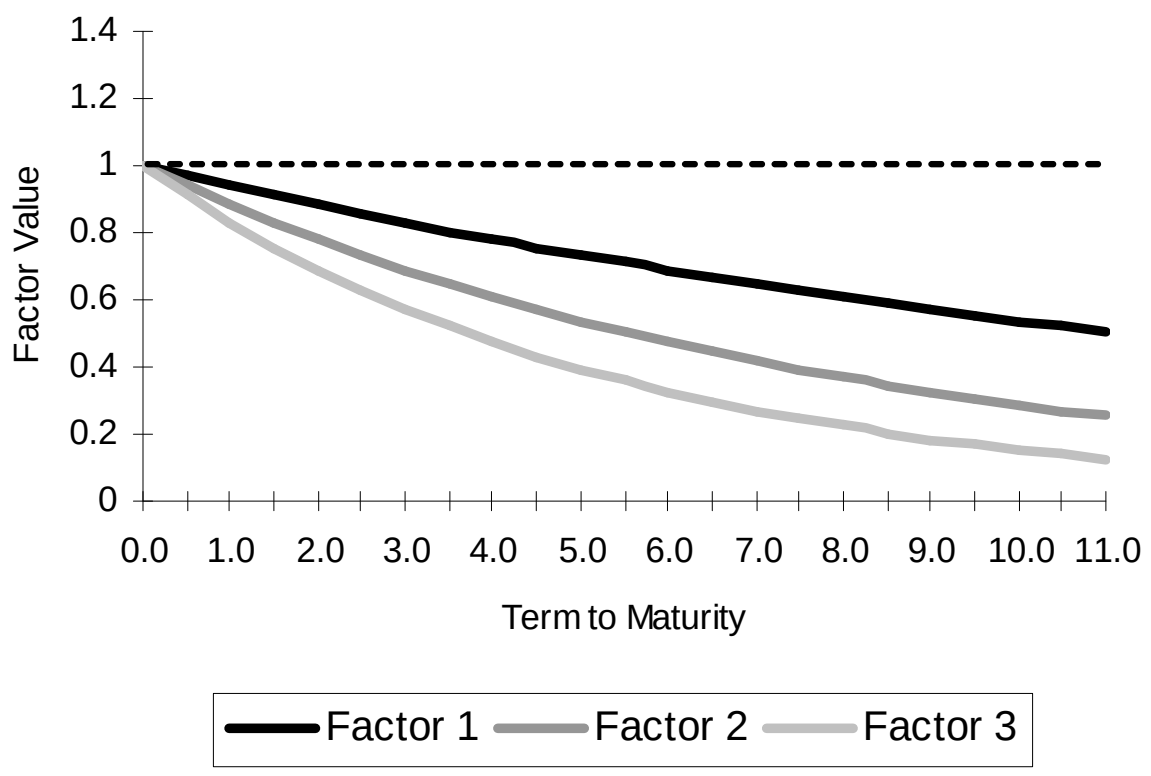

**Figure A2. Higher order (knot) spline factors**

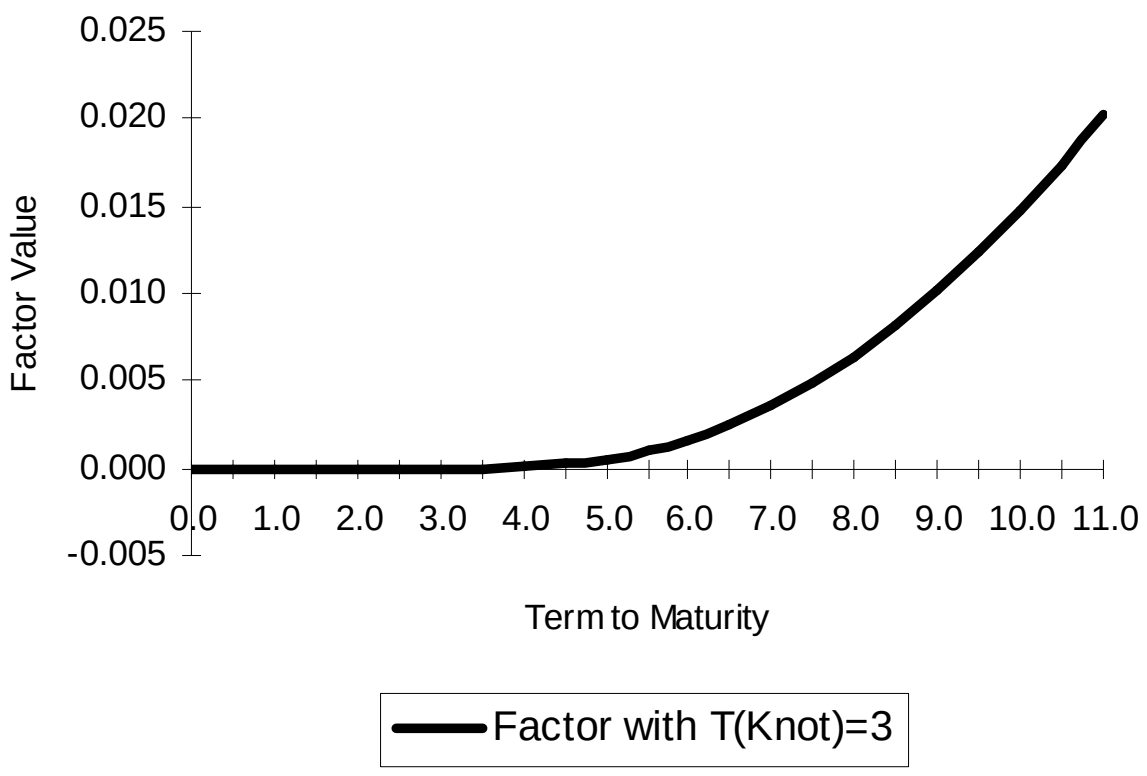